# A new key exchange cryptosystem

Li An-Ping

Beijing 100080, P.R. China
apli0001@sina.com

**Abstract:**

In this paper, we will present a new key exchange system based on linear algebra, which spend less operations but weaker in security than the Diffie-Hellman's one.

## 1. Introduction

The key exchange system is one of the main and most important public key cryptosystems, which was firstly proposed by W. Diffie and M.E. Hellman [1] based on discrete logarithm problem (DLP) over a finite group of large order. In which the two users of the system take $g^{k_1 \cdot k_2}$ as the communication key between them, where $g^{k_1}$ and $g^{k_2}$ are the public keys of them respectively.

There are an intensive researches on this system and discrete logarithm problem, for the detail material, refer to see [2].

In this paper, we will present a new key exchange system based on linear algebra, which will spend less operations but weaker in the security than the one of W. Diffie and M.E. Hellman.

## 2. Constructions and analysis

The mechanism of our new key exchange system is actually based on a simple observation in linear algebra.

Let $\mathbb{K}$ be a finite field $GF(q)$, and for a positive integer $m$, denoted by $V = \mathbb{K}^m$ the $m$-dimension vector space, and $M(m, \mathbb{K})$ the set of all $m \times m$ matrices over the finite field $\mathbb{K}$.

Let $\zeta \in V$, suppose that the two users A and B each takes a matrix $T_A, T_B \in M(k, \mathbb{K})$ as their privacy keys, and $T_A(\zeta)$, $T_B(\zeta)$ as their public keys respectively. If matrices $T_A$ and $T_B$ are commutative,

$$T_A \cdot T_B = T_B \cdot T_A, \qquad (2.1)$$

then the vector $T_A \cdot T_B(\zeta) = T_B \cdot T_A(\zeta)$ can be used as the communicative key between the user A and the user B.

Maybe, the easiest way to construct the privacy key matrices is that take privacy key matrices as

$$T = \sum_i c_i A^i, \quad c_i \in \mathbb{K}, \qquad (2.2)$$

where $A$ is a matrix with constant entries. However, it is clear that for the cryptographic applications it should prevent the privacy key $T$ from to be recovered from the public key $T(\zeta)$, and so each privacy key $T$ should have suffice many variable entries. For example, it is easy to know that the privacy keys constructed as the form (2.2) will be easy recovered when the matrix $A$ is known about its Jordan's decomposition.

In the following, we firstly describe our construction for the privacy keys and then provide a cryptanalysis for the new system.

Let $m = 2k$, denoted by $I$ the unit matrix of order $k$, and $\Gamma = \{\mu I \mid \mu \in \mathbb{K}\}$, and $\Lambda$ the set of $k \times k$ matrices with the form $\begin{pmatrix} \lambda & 1 & & \\ & \ddots & \ddots & \\ & & & 1 \\ & & & \lambda \end{pmatrix}$, $\lambda \in \mathbb{K}$, and $\Delta = \Gamma \cup \Lambda$. Clearly, the matrices in $\Delta$ are commutative. Moreover, let

$$\mathcal{A} = \left\{ \begin{pmatrix} x & 0 \\ 0 & x \end{pmatrix} \mid x \in \Delta \right\}, \quad \mathcal{B} = \left\{ \begin{pmatrix} a & b \\ c & d \end{pmatrix} \mid a,b,c,d \in \Delta \right\}, \tag{2.3}$$

then it is easy to verify that for any $A \in \mathcal{A}, B \in \mathcal{B}$, it has that

$$A \cdot B = B \cdot A. \tag{2.4}$$

Moreover, denoted by $\mathcal{Q}(\mathcal{A})$ and $\mathcal{Q}(\mathcal{B})$ the rings generated by $\mathcal{A}$ and $\mathcal{B}$ respectively. Then for any matrices $a \in \mathcal{Q}(\mathcal{A})$ and $z \in \mathcal{Q}(\mathcal{B})$, it has

$$z \cdot a = a \cdot z. \tag{2.5}$$

It should be noticed that in general the products of matrices in $\mathcal{B}$ are not commutative, so a mono-term in $\mathcal{Q}(\mathcal{B})$ is the form $B_1^{k_1} B_2^{k_2} \cdots B_s^{k_s}$, $B_i \in \mathcal{B}$, $k_i \geq 0$, $1 \leq i \leq s$.

Furthermore, for $z \in \mathcal{Q}(\mathcal{B})$, denoted by $\mathcal{Q}[z]$ the polynomial ring

$$\mathcal{Q}[z] = \left\{ \sum_i a_i z^i \mid a_i \in \mathcal{Q}(\mathcal{A}) \right\}. \tag{2.6}$$

Then for any $T_1, T_2 \in \mathcal{Q}[z]$, clearly,

$$T_1 \cdot T_2 = T_2 \cdot T_1. \tag{2.7}$$

That is, $\mathcal{Q}[z]$ is a commutative ring.

Thereby, the elements of $\mathcal{Q}[z]$ will be taken as the privacy keys.

Next, we give some cryptanalysis for the system presented.

Suppose that $T(\zeta) = \xi$, and $T_A$ is another privacy key, $T_A(\zeta) = \xi_A$, then it has

$$T_A(\xi) = T_A(T(\zeta)) = T(T_A(\zeta)) = T(\xi_A). \tag{2.8}$$

Let $T_A(\xi) = \rho_A$, it follows

$$T(\xi_A) = \rho_A. \tag{2.9}$$

Therefore, if there are $m$ linearly independent public keys $\xi_i = T_i(\zeta)$, $1 \leq i \leq m$, then the privacy key $T$ will be recovered.

*Note*. In fact, in general, the number of different variable entries of a matrix in $\mathcal{Q}[z]$ is equal to $2m$, so it is likely that the number of the equations as (2.9) required to recovery privacy key $T$ will be less than $m$. Of course, we may take the number of sub-matrices in $\mathcal{A}$ more than two, that is, take $m = dk$, $d \geq 2$. In this way, the number of different variable entries of a privacy key matrices in general is equal to $dm$.

Moreover, denoted by $\Omega$ the set of all the public keys, suppose that $rank(\Omega) = s$, and a adversary has $s$ pairs of privacy keys and public keys $(T_i, T_i(\zeta))$, $1 \leq i \leq s$, and $\{T_i(\zeta)\}_1^s$ are linearly independent, then the adversary will be able to recovery any communicative key of any privacy key $T$ without know the privacy key $T$.

Suppose that $\beta$ is a public key, then it can be written as

$$\beta = \sum_{1 \leq i \leq s} c_i T_i(\zeta). \tag{2.10}$$

So, the communicative key $T(\beta)$ for the privacy key $T$ can be represented as

$$T(\beta) = T(\sum_{1 \leq i \leq s} c_i T_i(\zeta)) = \sum_{1 \leq i \leq s} c_i T_i(T(\zeta)). \tag{2.11}$$

However, the communicative keys $T_i(T(\zeta))$, $1 \leq i \leq s$, are known for the adversary, and so he will recovery the communicative key $T(\beta)$.

The analyses above have demonstrated that the new system will be insecure if ones are able to access the privacy keys, even including themselves privacy keys. In other words, the privacy keys should be black for the users of the system.

## 2. Conclusion

The main advantage of the key exchange system presented is that spends less operations and so the implementation will be faster than Diffie-Hellman's one, for the process of key exchange here is only a linear transformation. But, we also have seen that the new system is weaker in security than Diffie-Hellman's one, and the applications will be restricted in the situations where the privacy keys are un-visible for the users, including himself privacy key, e.g. the communications with hardware.